\documentclass[english,twocolumn]{revtex4}
\usepackage[T1]{fontenc}
\usepackage[latin1]{inputenc}
\usepackage{graphicx}
\usepackage{amssymb}

\makeatletter

\renewcommand{\vec}[1]{{\mathbf #1}}
\newcommand{\bsigma}{\mbox{\boldmath{$\sigma$}}}
\newcommand{\bomega}{\mbox{\boldmath{$\omega$}}}

\newcommand{\qdotY}{2.0}  %orig: 3.6
\newcommand{\qdot}[1]{\begin{picture}(13,10)
    \put(6,\qdotY){\circle{13}}
    \put(6,\qdotY){\makebox(0,0){#1}}
    \end{picture}}

\newcommand{\qddot}[2]{\qdot{#1}$_{\!L}$\qdot{#2}$_{\!R}$}

\newcommand{\etal}[1]{ \textit{et al.}}

\newcommand{\comment}[1]{}

\usepackage{babel}
\makeatother
\begin{document}

\newcommand{\mubohr}{{\mu_{\mathrm{B}}}}

\newcommand{\efermi}{{\varepsilon_{\mathrm{F}}}}

\newcommand{\Dm}{\Delta\mu}

\newcommand{\bra}[1]{\langle#1|}

\newcommand{\ket}[1]{\left.|#1\rangle\right.}

\newcommand{\arrowSpace}{\hspace{-.75mm}}

\newcommand{\spup}{\ket{\arrowSpace\uparrow}}

\newcommand{\spdown}{\ket{ \arrowSpace\downarrow}}

\newcommand{\spupbra}{\bra{\uparrow\arrowSpace}}

\newcommand{\spdownbra}{\bra{\downarrow\arrowSpace}}

\newcommand{\spupup}{\ket{\arrowSpace\uparrow\uparrow}}

\newcommand{\spupdown}{\ket{\arrowSpace\uparrow\downarrow}}

\newcommand{\spdownup}{\ket{\arrowSpace\downarrow\uparrow}}

\newcommand{\spdowndown}{\ket{\! \downarrow\downarrow}}

\newcommand{\du}{{\downarrow\uparrow}}

\newcommand{\ud}{{\uparrow\downarrow}}

\newcommand{\upup}{{\uparrow\uparrow}}

\newcommand{\downdown}{{\downarrow\downarrow}}

\newcommand{\lblUpUp}{\uparrow\uparrow}

\newcommand{\lblUpDown}{\uparrow\downarrow}

\newcommand{\lblDownUp}{\downarrow\uparrow}

\newcommand{\lblDownDown}{\downarrow\downarrow}

\newcommand{\tQPC}{t^{\lblQPCleads}}

\newcommand{\VQPClr}{\Gamma_{\lblQPCleads}}

\newcommand{\decohRLintrinsic}{\Gamma_{\mathrm{i}}}

\newcommand{\decohRLtot}{\Gamma_{\mathrm{tot}}}

\newcommand{\suprDecohIntrinsic}{\eta}

\newcommand{\facCurrentSmallDm}{\lambda}

\newcommand{\VQPCpm}{V_{\mathrm{Q}}}

\newcommand{\VQPCup}{V_{\mathrm{Q}}^{\uparrow}}

\newcommand{\VQPCdown}{V_{\mathrm{Q}}^{\downarrow}}

\newcommand{\decohIntrinsic}{V_{-+}}

\newcommand{\chargeDecay}[1]{W_{#1}}

\newcommand{\ixDD}{\mathrm{d}}

\newcommand{\tDD}{t_{\ixDD}}

\newcommand{\Hdot}{H_{\mathrm{d}}}

\newcommand{\HDD}{V_{\mathrm{\ixDD}}}

\newcommand{\VdotElectrostatic}{V_{\mathrm{dot}}}

\newcommand{\ixQPCleads}{\mathrm{QPC}}

\newcommand{\lblQPCleads}{\mathrm{Q}}

\newcommand{\DE}{\varepsilon}

\newcommand{\DeltaEDown}{\DE_{\downarrow}}

\newcommand{\DeltaEUp}{\DE_{\uparrow}}

\newcommand{\DeltaEPlusMinus}{E}

\newcommand{\ZeemanRef}{\Delta_{z}^{\mathrm{r}}}

\newcommand{\ZeemanQubit}{\Delta_{z}^{\mathrm{q}}}

\newcommand{\efficiencyMeas}{e}

\newcommand{\visibilityRabi}{v}

\newcommand{\binomProb}[1]{p_{#1}}

\newcommand{\lblL}{L}

\newcommand{\lblR}{R}

\newcommand{\ixLeftDot}{\lblL}

\newcommand{\ixRightDot}{\lblR}

\newcommand{\ixTripP}{T_{+}}

\newcommand{\ixTripM}{T_{-}}

\newcommand{\ixTripZero}{T_{0}}

\newcommand{\rhoStatTripP}{|T_{+}\rangle\langle T_{+}|}

\newcommand{\rhoStatTripZero}{|T_{0}\rangle\langle T_{0}|}

\newcommand{\lblSR}{S_{\lblR}}

\newcommand{\lblSLR}{S_{\lblL\lblR}}

\newcommand{\lblPsiP}{+}

\newcommand{\lblPsiM}{-}

\newcommand{\lblPsiPM}{\pm}

\newcommand{\ixPsiP}{+}

\newcommand{\ixPsiM}{-}

\newcommand{\ixPsiPM}{\pm}

\newcommand{\WPsiPm}{W_{Q}}

\newcommand{\fermiProdOverDmSymbol}{g}

\newcommand{\fermiProdOverDm}[1]{\fermiProdOverDmSymbol(#1)}

\newcommand{\fermiProdOverDmPMSum}{\fermiProdOverDmSymbol_{\Sigma}}

\newcommand{\tempDecohFac}{\kappa}

\newcommand{\BlochEqMSymbol}{P}

\newcommand{\BlochEqM}{\vec{\BlochEqMSymbol}}

\newcommand{\BlochIncohMatrix}{\Gamma}

\newcommand{\currentMxElemSq}[1]{\mathcal{I}_{#1}}

\newcommand{\measOutcome}[1]{A_{#1}}

\newcommand{\POVMop}[1]{E_{\measOutcome{#1}}}

\newcommand{\expectQubitMeas}[2]{{\langle#1\rangle}_{#2}}

\newcommand{\rhoFullIx}{\mathrm{f}}

\newcommand{\sysLbl}{S}

\newcommand{\rhoSys}{\rho}

\newcommand{\rSNot}{\rhoSys^{0}}

\newcommand{\rhoSysStat}{\bar{\rho}}

\newcommand{\rBNot}{{\rho_{\lblQPCleads}^{0}}}

\newcommand{\LV}{L_{V}}

\newcommand{\TrB}{\mathrm{Tr}{}_{\lblQPCleads}}

\newcommand{\currentStat}{\bar{I}}

\newcommand{\currentStatIncoh}{\currentStat_{\mathrm{incoh}}}

\newcommand{\gammaOut}{\gamma_{\mathrm{out}}}

\newcommand{\gammaIn}{\gamma_{\mathrm{in}}}

\newcommand{\timeQPCmeas}{t_{\mathrm{m}}}

\title{Measurement efficiency and n-shot read out of spin qubits }

\author{Hans-Andreas Engel}

\author{Vitaly Golovach}

\author{Daniel Loss}

\affiliation{Department of Physics and Astronomy, University of Basel, Klingelbergstrasse
82, CH-4056 Basel, Switzerland}

\author{L.M.K. Vandersypen}

\author{J.M. Elzerman}

\author{R. Hanson}

\author{L.P. Kouwenhoven}

\affiliation{Department of NanoScience and ERATO Mesoscopic Correlation Project,
PO Box 5046, 2600 GA Delft, The Netherlands}

\begin{abstract}
We consider electron spin qubits in quantum dots and define a measurement
efficiency $\efficiencyMeas$ to characterize reliable measurements
via $n$-shot read outs. We propose various implementations based
on a double dot and quantum point contact (QPC) and show that the
associated efficiencies $\efficiencyMeas$ vary between 50\% and 100\%,
allowing single-shot read out in the latter case. We model the read
out microscopically and derive its time dynamics in terms of a generalized
master equation, calculate the QPC current and show that it allows
spin read out under realistic conditions.

\end{abstract}
\maketitle

The read out of a qubit state is of central importance for quantum
information processing \cite{NielsenChuang}. In special cases, the
qubit state can be determined in a single measurement, referred to
as single shot read out. In general, however, the measurement needs
to be performed not only once but $n$ times, where $n$ depends on
the qubit, the efficiency $\efficiencyMeas$ of the measurement device,
and on the tolerated inaccuracy (infidelity) $\alpha$. In the first
part of this Letter, we analyze such $n$-shot read outs for general
qubit implementations and derive a lower bound on $n$ in terms of
$\efficiencyMeas$ and $\alpha$. We then turn to spin-based qubits
and GaAs quantum dots \cite{Loss97,reviewQdot} and analyze their
$n$-shot read out based on a spin-charge conversion and charge measurement
via quantum point contacts.

\emph{$n$-shot read out and measurement efficiency $\efficiencyMeas$.}
How many times $n$ do the preparation and measurement need to be
performed until the state of the qubit is known with some given infidelity
$\alpha$ ($n$-shot read out)? We consider a well-defined qubit,
i.e., we take only a two-dimensional qubit Hilbert space into account
and exclude leakage to other degrees of freedom. We define a set
of positive operator-valued measure (POVM) operators \cite{Peres},
$\POVMop{0}=p_{0}\ket{0}\bra{0}+(1-p_{1})\ket{1}\bra{1}$ and $\POVMop{1}=(1-p_{0})\ket{0}\bra{0}+p_{1}\ket{1}\bra{1}$,
where $p_{0}$ and $p_{1}$ are probabilities. These operators describe
measurements with outcomes $\measOutcome{0}$ and $\measOutcome{1}$,
\emph{resp}. They are positive and $\POVMop{0}+\POVMop{1}=1$. This
model of the measurement process can be pictured as follows. First,
the qubit is coupled to some other device (e.g., to a reference dot,
see below). Then this coupled system is measured and thereby projected
onto some internal state. That state is accessed via an external {}``pointer''
observable $\hat{A}$ \cite{Peres} (e.g., a particular charge distribution,
a time-averaged current, or noise). We assume that only two measurement
outcomes are possible, either $\measOutcome{0}$ or $\measOutcome{1}$,
which are classically distinguishable \cite{fnUncertOutcome}.  For
initial qubit state $\ket{0}$ the expectation value is $\expectQubitMeas{\hat{A}}{0}=p_{0}\measOutcome{0}+(1-p_{0})\measOutcome{1}$,
while for initial state $\ket{1}$ it is $\expectQubitMeas{\hat{A}}{1}=(1-p_{1})\measOutcome{0}+p_{1}\measOutcome{1}$.
Let us take an initial qubit state $\ket{0}$and consider a single
measurement. With probability $p_{0}$, the measurement outcome is
$\measOutcome{0}$ which one would interpret as {}``qubit was in
state $\ket{0}$''. However, with probability $1-p_{0}$, the outcome
is $\measOutcome{1}$ and one might incorrectly conclude that {}``qubit
was in state $\ket{1}$''. Conversely, the initial state $\ket{1}$
leads with probability $p_{1}$ to $\measOutcome{1}$ and with $1-p_{1}$
to $\measOutcome{0}$. We now determine $n$ for a given $\alpha$,
for a qubit either in state $\ket{0}$ \emph{or} $\ket{1}$ (no superposition
allowed \cite{effSuperpos}). For an accurate read out we need, roughly
speaking, that $\expectQubitMeas{\hat{A}}{0}$ and $\expectQubitMeas{\hat{A}}{1}$
are separated by more than the sum of the corresponding standard deviations.
More precisely \cite{Bosch}, we consider a parameter test of a binomial
distribution of the measurement outcomes, one of which is $\measOutcome{0}$
with probability $p$. The null hypothesis is that the qubit is in
state $\ket{0}$, thus $p=p_{0}$. The alternative is a qubit in state
$\ket{1}$, thus $p=1-p_{1}$. For sufficiently large $n$, namely
$n\, p_{0,1}(1-p_{0,1})>9$, one can approximate the binomial with
a normal distribution \cite{footnoteExpG}. The state of the qubit
can then be determined with significance level \emph{}({}``infidelity'')
$\alpha$ for \begin{eqnarray}
n & \geq & z_{1-\alpha}^{2}\Big(\frac{1}{e}-1\Big),\label{eqnnMin}\\
\efficiencyMeas & = & \left(\sqrt{\binomProb{0}\binomProb{1}}-\sqrt{(1-\binomProb{0})(1-\binomProb{1})}\right)^{2},\label{eqnMEffG}\end{eqnarray}
 with the quantile (critical value) $z_{1-\alpha}$ of the standard
normal distribution function, $\Phi(z_{1-\alpha})=1-\alpha=\frac{1}{2}\big[1+\mathrm{erf}(z_{1-\alpha}/\sqrt{2})\big]$.
We interpret $\efficiencyMeas$ as \emph{measurement efficiency}.
Indeed, it is a single parameter $\efficiencyMeas\in[0,\,1]$ which
tells us if $n$-shot read out is possible. For $p_{0}=p_{1}=1$,
the efficiency is maximal, $\efficiencyMeas=100\%$, and single-shot
read out is possible ($n=1$). Conversely, for $p_{1}=1-p_{0}$ (e.g.,
$p_{0}=p_{1}=\frac{1}{2}$), the state of the qubit cannot be determined,
not even for an arbitrarily large $n$, and the efficiency is $\efficiencyMeas=0\%$.
For the intermediate regime, $0\%<\efficiencyMeas<100\%$, the state
of the qubit is known after several measurements, with $n$ satisfying
Eq.~(\ref{eqnnMin}).

\emph{Visibility} $\visibilityRabi$. When coherent oscillations between
$\ket{0}$ and $\ket{1}$ are considered, the amplitude of the oscillating
signal is $\big|\expectQubitMeas{\hat{A}}{1}-\expectQubitMeas{\hat{A}}{0}\big|$,
i.e., smaller than the value $\left|\measOutcome{1}-\measOutcome{0}\right|$
by a factor of $\visibilityRabi=\left|p_{0}+p_{1}-1\right|.$ Thus,
we can take $\visibilityRabi$ as a measure of the visibility of the
coherent oscillations. With $\visibilityRabi$ and the shift of the
oscillations, $s=\frac{1}{2}\left(p_{1}-p_{0}\right)=\frac{1}{2}\big(\expectQubitMeas{\hat{A}}{0}+\expectQubitMeas{\hat{A}}{1}-\measOutcome{0}-\measOutcome{1}\big)/\big(\measOutcome{1}-\measOutcome{0}\big)$,
we can get $\efficiencyMeas$. We find the general relation $\visibilityRabi^{2}\leq\efficiencyMeas\leq\visibilityRabi$,
where the left inequality becomes exact for $p_{0}=p_{1}$ and the
right for $p_{0}=1$ or $p_{1}=1$. Further, for every $0<\epsilon<1$
we can take $p_{0}=\frac{1}{2}$ and $p_{1}=\frac{1}{2}+\frac{\epsilon}{2}$,
thus $\efficiencyMeas<\epsilon\visibilityRabi$. Hence, given these
natural interpretations of $\efficiencyMeas$ and $\visibilityRabi$,
we see that somewhat unexpectedly the efficiency can be much smaller
than the visibility (of course, $\efficiencyMeas=0\Leftrightarrow\visibilityRabi=0$).

\begin{figure}
\includegraphics[%
  width=40mm]{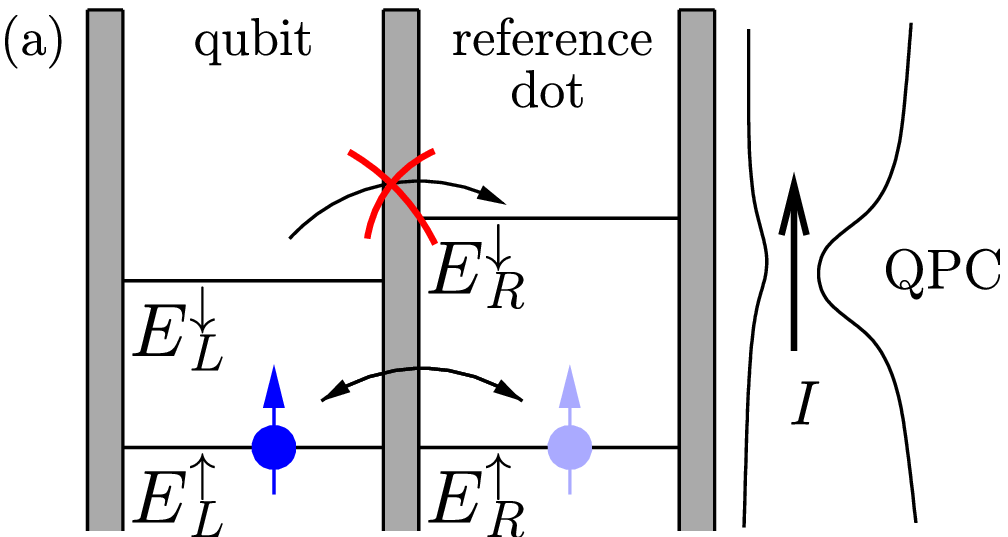}~~~~~\includegraphics[%
  width=40mm]{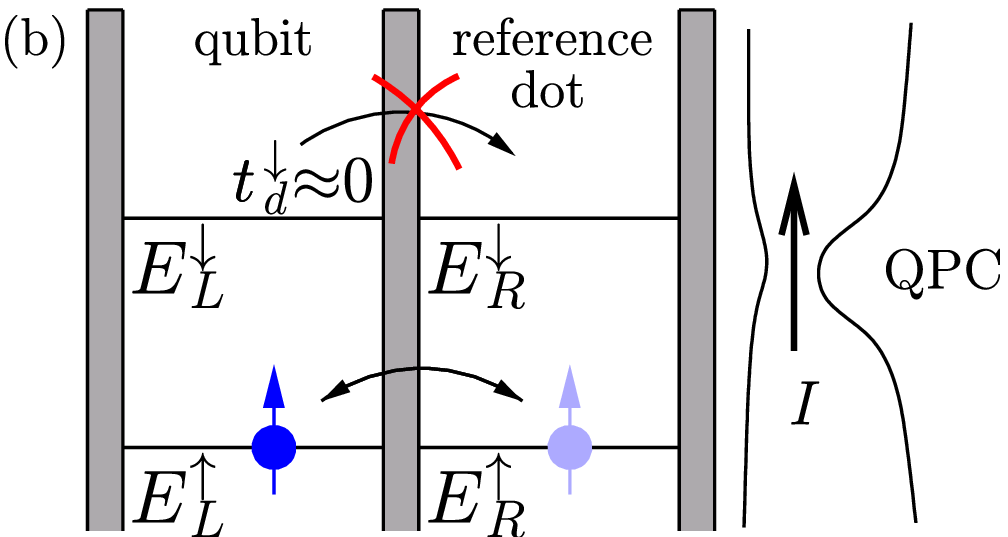}

\begin{flushleft}\includegraphics[%
  width=40mm]{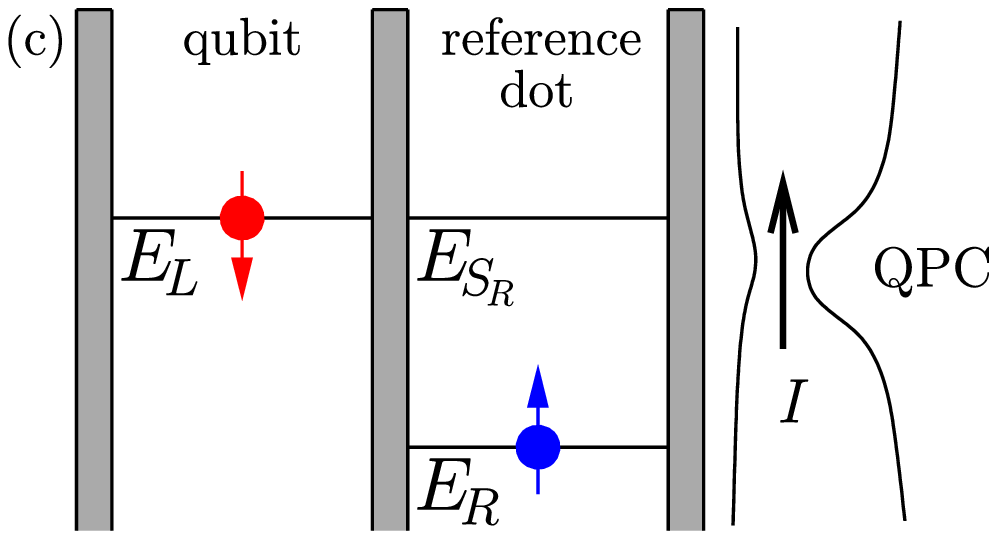}~~~~~~~\includegraphics[%
  bb=0bp 55bp 288bp 247bp,
  width=31mm]{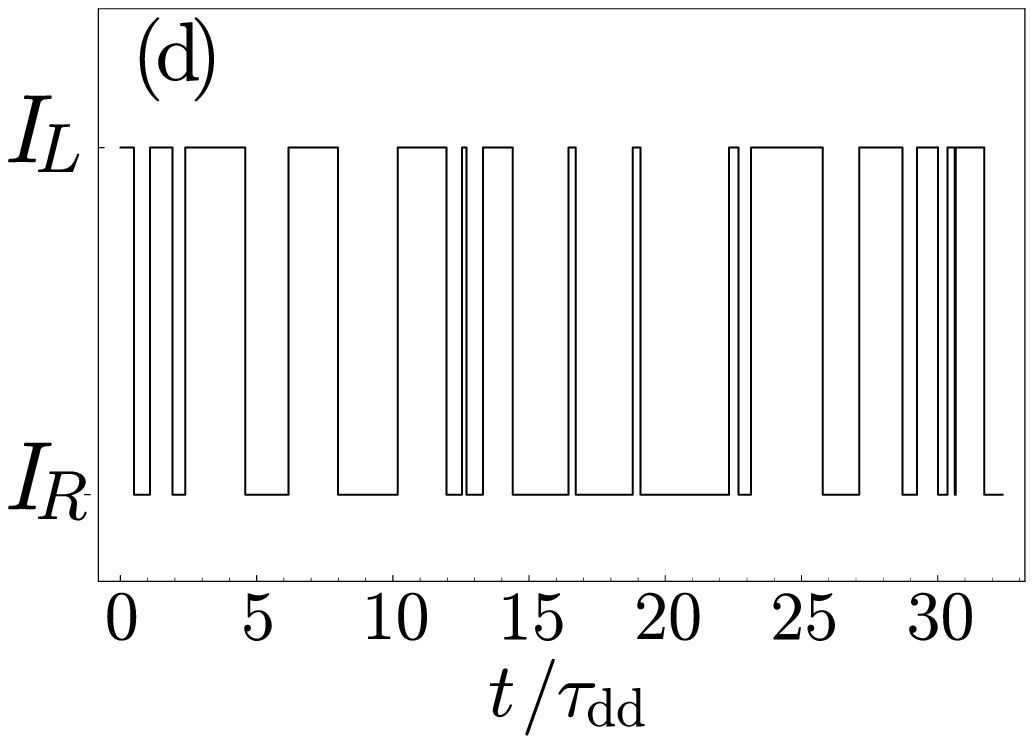}\end{flushleft}

\caption{\label{figROdd}Electron spin read-out setup consisting of a double
dot. The right {}``reference'' dot is coupled capacitively to a
QPC shown on the right. (a) Read out using different Zeeman splittings.
For $\uparrow$, the electron tunnels between the two dots. For $\downarrow$,
tunneling is suppressed by the detuning and the stationary state has
a large contribution of the left dot since it has lower energy. This
allows single-shot read out, i.e., $\efficiencyMeas=100\%$. (b) Spin-dependent
tunneling amplitudes, $\tDD^{\downarrow}<\tDD^{\uparrow}$, also enable
efficient read out. (c) Read out with the singlet state. Tunneling
of spin $\uparrow$ to the reference dot is blocked due to the Pauli
principle. (d) \label{figTelegraph}Schematic current vs.\ time during
a single measurement. Here, $\tau_{\mathrm{dd}}$ is the time scale
for tunneling and we assume $\decohRLtot>\tDD$, i.e., that the tunneling
events can be resolved in the current.}
\end{figure}
\emph{Single spin read out}. We now discuss several concrete read-out
setups and their measurement efficiency. We consider a promising qubit,
which is an electron spin confined in a quantum dot \cite{Loss97,reviewQdot}.
For the read out of such a spin qubit, the time scale is limited by
the spin-flip time $T_{1}$, which  has a lower bound of $\approx100\:\mu\mathrm{s}$
\cite{FujisawaT1,HansonT1} (while $T_{2}$ is not of relevance here).
One setup proposed in Ref.\ \onlinecite{Loss97} is read out via
a neighboring paramagnetic dot, where the qubit spin nucleates formation
of a ferromagnetic domain. This leads to $p_{0}=p_{1}=\frac{3}{4}$
and thus $\efficiencyMeas=25\%$. Another idea is to transfer the
qubit information from spin to charge \cite{Loss97,reviewQdot,KaneSET,Recher,ELesr}.
For this, we propose to couple the qubit dot to a second ({}``reference'')
dot \cite{footnoteRefLead} and discuss several possibilities how
that coupling can be made spin-dependent, see also Fig~\ref{figROdd}.
 The resulting charge distribution on the double dot will then depend
on the qubit spin state and can be detected by coupling the double
dot to an electrometer, such as a quantum point contact (QPC) \cite{Field,Elzerman},
see Fig~\ref{figROdd} (or, alternatively, a single-electron transistor
\cite{Rimberg}).

\emph{Read out with different Zeeman splittings}. First, we propose
a setup where efficiencies up to $100\,\%$ can be reached, see Fig.~\ref{figROdd}a.
We take a double dot with different Zeeman splittings, $\Delta_{z}^{\ixLeftDot,\ixRightDot}=E_{\ixLeftDot,\ixRightDot}^{\downarrow}-E_{\ixLeftDot,\ixRightDot}^{\uparrow}$,
in each dot \cite{footnoteInhZeeman} and consider a single electron
on the double dot.  For initial qubit state $\spup$, the electron
can tunnel from state $\ket{\lblL_{\uparrow}}\widehat{=}$ \qddot{$\uparrow$}{$$}
to state $\ket{\lblR_{\uparrow}}\widehat{=}$ \qddot{$$}{$\uparrow$}
and vice versa, and analogously for qubit state $\spdown$. We consider
time scales shorter than $T_{1}$, thus the states with different
spins are not coupled. Next, we define the detunings $\DE_{\uparrow,\downarrow}=E_{\ixLeftDot}^{\uparrow,\downarrow}-E_{\ixRightDot}^{\uparrow,\downarrow}$,
which are different for the up and down states, $\DeltaEDown-\DeltaEUp=\Delta_{z}^{\ixLeftDot}-\Delta_{z}^{\ixRightDot}\neq0$.
The stationary state of the double dot depends on $\DE_{\uparrow,\downarrow}$
and so does the QPC current $\bar{I}_{\uparrow,\downarrow}$  {[}we
show this below, see Eq. (\ref{eqnCurrentStat}) and $\currentStatIncoh${]}.
Therefore, initial states $\spup$ and $\spdown$ can be identified
through distinguishable stationary currents \cite{fnUncertOutcome},
$\currentStat_{\uparrow}\neq\currentStat_{\downarrow},$ \emph{thus
$\efficiencyMeas={}$100\% and single-shot read out is possible. }

\emph{Spin-dependent tunneling} provides another read-out scheme,
see Fig.~\ref{figROdd}b, which we describe with spin-dependent tunneling
amplitudes $\tDD^{\uparrow,\downarrow}$. For $\tDD^{\downarrow}\ll\tDD^{\uparrow}$,
only spin $\uparrow$ tunnels onto the reference dot while tunneling
of spin $\downarrow$ is suppressed. We assume the same Zeeman splitting
in both dots and resonance $\DE=0$. It turns out {[}Eq. (\ref{eqnCurrentStat}){]}
that $\bar{I}_{\uparrow,\downarrow}$ depends on $\tDD^{\uparrow,\downarrow}$
and thus the state of the qubit can be measured.  However, the decay
to the stationary state is quite slow in case the qubit is $\spdown$,
due to the suppressed tunneling amplitude $\tDD^{\downarrow}$ . Since
the difference in charge distribution between qubit $\spup$ and $\spdown$
is larger at short timescales, it can thus be advantageous to measure
the time-dependent current (discussed toward the end).

\emph{Read out with Pauli principle}. We now consider the case where
the reference dot contains initially an electron in spin up ground
state, see Fig.~\ref{figROdd}c. We assume gate voltages such that
there are either two electrons on the right dot or one electron on
each dot. Thus, we consider the 5 dimensional Hilbert space $\ket{{\lblSR}}\widehat{=}$
\setlength{\unitlength}{0.95pt} \qddot{$$}{$$$\ud$}, $\spupdown\widehat{=}$
\qddot{$\uparrow$}{$\downarrow$}, $\spdownup\widehat{=}$ \qddot{$\downarrow$}{$\uparrow$},
$\ket{T_{+}}\widehat{=}$ \qddot{$\uparrow$}{$\uparrow$}, $\ket{T_{-}}\widehat{=}$
\qddot{$\downarrow$}{$\downarrow$}. We define the {}``delocalized''
singlet $\ket{\lblSLR}=(\spupdown-\spdownup)/\sqrt{2}$ and the triplet
$\ket{T_{0}}=(\spupdown+\spdownup)/\sqrt{2}$. In the absence of tunneling,
the corresponding energies are $E_{\lblSR}=2\epsilon_{R}+U$ and $E_{\lblSLR}=E_{T_{0,\pm}}=\epsilon_{L}+\epsilon_{R}$
with charging energy $U$ and single particle energies $\epsilon_{L,R}$.
We can neglect states with two electrons on the qubit dot and the
triplet states with two electrons on the reference dot, since they
have a much larger energy (their admixture due to tunneling is small).
We denote the state with an {}``extra'' electron on the right dot
as $\ket{\lblR}\equiv\ket{\lblSR}$ with corresponding QPC current
$I_{\ixRightDot}$. For state $\ket{\lblL}\equiv\ket{\lblSLR}$ and
for all triplet states, $\ket{T_{0,\pm}}$, the current is $I_{\ixLeftDot}$.
When tunneling is switched on and the qubit is initially in state
$\spup$, tunneling to the reference dot is blocked due to the Pauli
exclusion principle \cite{OnoPauli}. Thus, the double dot will remain
in the (stationary) state $\rhoStatTripP$ and the current in the
quantum dot remains $\langle I\rangle=I_{\ixLeftDot}$ (a so-called
non-demolition measurement). On the other hand, for an initial qubit
state $\spdown$, the initial state of the double dot is $\ket{\du}=(\ket{T_{0}}-\ket{\lblSLR})/\sqrt{2}$.
The contribution $\ket{\lblSLR}$ of this superposition is tunnel
coupled to $\ket{\lblSR}$ and will decay to the stationary state
$\rhoSysStat$ with corresponding QPC current $\currentStat$ (see
below for an explicit evaluation). In contrast, the triplet contribution
$\ket{T_{0}}$ is not tunnel-coupled to $\ket{\lblSR}$ due to spin
conservation and does not decay. In total, the density matrix of the
double dot decays into the stationary value $\frac{1}{2}(\rhoStatTripZero+\rhoSysStat)$.
For $\DE=0$, the ensemble-averaged QPC current for qubit $\spdown$
is $\langle I\rangle=\frac{1}{2}(I_{\ixLeftDot}+\currentStat)\approx\frac{1}{4}(3I_{\ixLeftDot}+I_{\ixRightDot})$
and can thus be distinguished from $I_{\ixLeftDot}$ for qubit $\spup$.
 However, in a single run of such a measurement, an initial qubit
$\spdown$ decays either into $\rhoStatTripZero$ or into $\rhoSysStat$,
with 50\% probability each. Since $\rhoStatTripZero$ and $\rhoStatTripP$
lead to the same QPC current $I_{\ixLeftDot}$, these two states are
not distinguishable within this read-out scheme and single-shot read-out
is not possible. The read out can now be described with the POVM model
given above, with $\spup\equiv\ket{0}$ and $\spdown\equiv\ket{1}$
and $\measOutcome{\uparrow}=I_{\ixLeftDot}$; $\measOutcome{\downarrow}=\currentStat$;
$p_{\uparrow}=1$; and $p_{\downarrow}=\frac{1}{2}$. Thus, the \emph{measurement
efficiency is $\efficiencyMeas={}$50\%,} i.e., to achieve a fidelity
of $1-\alpha=99\%$, we need $n\geq7$ read outs \cite{footnoteExpG}.

An analogous read out is possible if the ground state of the reference
dot is a triplet, say $\ket{RT_{+}}\widehat{=}$ \qddot{$$}{$\uparrow\uparrow$}
which is lower than the other triplets ($\ket{RT_{0,-}}$, $\ket{RT_{-}}$)
due to Zeeman splitting. Again, we assume that the reference dot is
initially $\spup$. First, for a qubit state $\spup$ and at resonance,
$\DE=0$, tunneling into $\ket{RT_{+}}$ always occurs and $p_{\uparrow}=1$.
Second, the qubit state $\spdown$ has an increased energy by the
Zeeman splitting $\Delta_{z}$ and is thus at resonance with $\ket{RT_{0}}$
(which has also an increased energy). If the double dot is not projected
onto the singlet (in $50\,\%$ of the cases), tunneling onto the reference
dot will also occur, i.e., $p_{\downarrow}=\frac{1}{2}$. Thus, when
one detects an additional charge on the reference dot, the initial
state of the qubit is not known. We find again $\efficiencyMeas=50\%$.

\emph{Read-out model.} So far we have introduced various spin read
out schemes and the corresponding measurement efficiencies. In order
to evaluate the signal strength $\measOutcome{0}-\measOutcome{1}$
for these schemes, we now calculate the stationary charge distribution
$\bar{\rho}$ and QPC current $\currentStat$ for the case when the
electron can tunnel coherently between the two dots (as a function
of the detuning and the tunnel coupling). We describe the read-out
setup with the Hamiltonian $H=\Hdot+\HDD+H_{\mathrm{\ixQPCleads}}+V.$
Here, $H_{\mathrm{\ixQPCleads}}$ contains the energies of the (uncoupled)
Fermi leads of the QPC. Further, $\Hdot$ describes the double dot
in the absence of tunneling, including orbital and electrostatic charging
energies, $\Hdot\ket{n}=E_{n}\ket{n}$. It thus contains $\DE=E_{\ixLeftDot}-E_{\ixRightDot}$,
the detuning of the tunneling resonance. The inter-dot tunneling Hamiltonian
is defined as $\HDD=\tDD(\ket{\lblR}\bra{\lblL}+\ket{\lblL}\bra{\lblR})$.
(Note that for tunneling between $\ket{\lblSLR}$ and $\ket{\lblSR}$,
$\tDD$ is $\sqrt{2}$ times the one-particle tunneling amplitude,
since both states $\ket{\ud}$ and $\ket{\du}$ are involved). $V$
is a tunneling Hamiltonian describing transport through the QPC. The
tunneling amplitudes, $\tQPC_{\ixLeftDot}$ and $\tQPC_{\ixRightDot}$,
will be influenced by electrostatic effects, in particular by the
charge distribution on the double dot. Thus, we model the measurement
of the dot state via the QPC with $V=\left(\tQPC_{\ixLeftDot}\ket{\lblL}\bra{\lblL}+\tQPC_{\ixRightDot}\ket{\lblR}\bra{\lblR}\right)\sum\big(c_{\mathrm{in}}^{\dagger}c_{\mathrm{out}}^{{\displaystyle {}}}+\mathrm{h.c.}\big)$
\cite{Gurvitz,Korotkov,Goan}. Here, $c_{\mathrm{in}}^{\dagger}$
and $c_{\mathrm{out}}^{\dagger}$ create electrons in the incoming
and the outgoing leads of the QPC, where the sum is taken over all
momentum and spin states. We derive the master equation for the reduced
density matrix $\rhoSys$ of the double dot. We use standard techniques
and make a Born-Markov approximation in $V$ \cite{Blum,footnoteTwoLevel}.
 We allow for an arbitrary inter-dot tunnel coupling, i.e., we keep
$\HDD$ exactly, with energy splitting $\DeltaEPlusMinus=\sqrt{4\,\tDD^{2}+\DE^{2}}$
in the eigenbasis of $\Hdot+\HDD$. We obtain the master equation~\cite{footnoteBloch}\begin{eqnarray}
{\dot{\rho}_{\ixLeftDot}} & = & -{\dot{\rho}_{\ixRightDot}}=2\tDD\,\mathrm{Im}\,[{\rho_{\ixRightDot\ixLeftDot}}],\label{eqnMasterL}\\
{\dot{\rho}_{\ixRightDot\ixLeftDot}} & = & \left[i\tDD+\tDD\frac{\VQPClr\DE}{\DeltaEPlusMinus^{2}}(\fermiProdOverDmPMSum-2\fermiProdOverDmSymbol_{0})\right]\left({\rho_{\ixRightDot}}-{\rho_{\ixLeftDot}}\right)\nonumber \\
 &  & -\frac{\tDD\,{\VQPClr}}{{\Dm}}-\left(\tempDecohFac{\VQPClr}+\decohRLintrinsic-i{\DE}\right){\rho_{\ixRightDot\ixLeftDot}},\label{eqnMasterRL}\end{eqnarray}
for $\rho_{n}=\bra{n}\rhoSys\ket{n}$ and $\rho_{\ixRightDot\ixLeftDot}=\bra{\lblR}\rhoSys\ket{\lblL}$.
In comparison to previous work \cite{Gurvitz,Korotkov,Goan}, we find
an additional term, $-\tDD\,{\VQPClr}/\Dm$, which comes from treating
$\HDD$ exactly. We find that the current through the QPC is $I_{\ixLeftDot}=2\pi\nu^{2}e\,\Dm|\tQPC_{\ixLeftDot}|^{2}$
for state $\ket{\lblL}$ and analogously $I_{\ixRightDot}$ for state
$\ket{\lblR}$, and we choose $I_{\ixLeftDot},\, I_{\ixRightDot}\geq0$.
Here, $\Dm>0$ is the applied bias across the QPC and $\nu$ is the
DOS at the Fermi energy of the leads connecting to the QPC. We define
$\fermiProdOverDmSymbol_{\pm}=\fermiProdOverDm{\Dm\pm\DeltaEPlusMinus}$,
$\fermiProdOverDmPMSum=\fermiProdOverDmSymbol_{+}+\fermiProdOverDmSymbol_{-}$
and $\fermiProdOverDmSymbol_{0}=\fermiProdOverDm{\Dm}$ with $\fermiProdOverDm{x}=x\big/\Dm\big(e^{x/kT}-1\big)$.
The values $\fermiProdOverDmSymbol_{\pm,\Sigma,0}$ vanish for $\Dm\pm\DeltaEPlusMinus>kT$.
In this case, the decay rate due to the current assumes the known
value \cite{Gurvitz,Korotkov,Goan}, $\VQPClr=\left(\sqrt{I_{\ixLeftDot}}-\sqrt{I_{\ixRightDot}}\right)^{2}\big/\,2e$.
Generally, the factor $\tempDecohFac=1+(4\tDD^{2}\fermiProdOverDmPMSum+2\DE^{2}\fermiProdOverDmSymbol_{0})/\DeltaEPlusMinus^{2}$
accounts for additional relaxation/dephasing due to particle hole
excitations, induced, e.g., by thermal fluctuations of the QPC current.
For almost equal currents, $I_{\ixLeftDot,\,\ixRightDot}=I\,(1\pm\frac{1}{2}x)$,
we have $\VQPClr=Ix^{2}/8e+O(x^{4})$. Finally, by introducing the
phenomenological rate $\decohRLintrinsic$ we have allowed for some
intrinsic charge dephasing, which occurs on the time scale of nanoseconds
\cite{FujisawaT2dd}. For an initial state in the subspace $\{\ket{\lblL},\,\ket{\lblR}\}$,
we find the stationary solution of the double dot, $\rhoSysStat=\frac{1}{2}(1-\suprDecohIntrinsic\DE/\Dm)\ket{\lblL}\bra{\lblL}+\frac{1}{2}(1+\suprDecohIntrinsic\DE/\Dm)\ket{\lblR}\bra{\lblR}-\suprDecohIntrinsic(\tDD/\Dm)(\ket{\lblR}\bra{\lblL}+\ket{\lblL}\bra{\lblR})$,
where $\suprDecohIntrinsic=\VQPClr/[\VQPClr(1+\fermiProdOverDmPMSum)+\decohRLintrinsic]$.
Positivity of $\rhoSysStat$ is satisfied since $\suprDecohIntrinsic\leq\Dm/\DeltaEPlusMinus$.
  The time decay to $\rhoSysStat$ is described by three rates,
given as the roots of $P(\lambda)=\lambda^{3}+2\decohRLtot\lambda^{2}+\left(\DeltaEPlusMinus^{2}+\decohRLtot^{2}\right)\lambda+4\tDD^{2}\big[\decohRLtot+\VQPClr(\fermiProdOverDmPMSum-2\fermiProdOverDmSymbol_{0})\DE^{2}/\DeltaEPlusMinus^{2}\big]$,
with $\decohRLtot=\tempDecohFac{\VQPClr}+\decohRLintrinsic$. The
stationary current through the QPC is given by $\bar{I}=\rhoSysStat_{\ixLeftDot}I_{\ixLeftDot}+\rhoSysStat_{\ixRightDot}I_{\ixRightDot}+2e\,\tDD\facCurrentSmallDm(\VQPClr/\Dm)\,\mathrm{Re\,\rhoSysStat_{\ixRightDot\ixLeftDot}}$
and thus becomes\begin{equation}
\currentStat=\frac{I_{\ixLeftDot}+I_{\ixRightDot}}{2}+\eta\frac{\DE}{2\Dm}\left(I_{\ixRightDot}-I_{\ixLeftDot}\right)-\eta\facCurrentSmallDm\frac{2e\,\VQPClr\tDD^{2}}{\Dm^{2}},\label{eqnCurrentStat}\end{equation}
 where $\facCurrentSmallDm=1-\Dm(\fermiProdOverDmSymbol_{-}-g_{+})/\DeltaEPlusMinus$.
We note that $\eta$ quantifies the effect of the detuning $\DE$
on the QPC current. To reach maximal sensitivity, $\eta=1$, we need
$I_{\ixRightDot}\lesssim I_{\ixLeftDot}/10$ for $I\sim1\,\mathrm{nA}$
and $\decohRLintrinsic\sim10^{9}\:\mathrm{s}^{-1}$. In linear response,
the current becomes $(I_{\ixLeftDot}+I_{\ixRightDot})/2+\left(I_{\ixRightDot}-I_{\ixLeftDot}\right)\DE\tanh(\DeltaEPlusMinus/2kT)[1-(\Gamma_{i}\Dm/\VQPClr\DeltaEPlusMinus)\tanh(\DeltaEPlusMinus/2kT)]/2\DeltaEPlusMinus-2e\,\tDD^{2}\VQPClr[1-\DeltaEPlusMinus/kT$
$\sinh(\DeltaEPlusMinus/kT)]/\DeltaEPlusMinus^{2}+e\,\tDD^{2}\Gamma_{i}\Dm[\sinh(\DeltaEPlusMinus/kT)-\DeltaEPlusMinus/kT][1-\Gamma_{i}\Dm$
$\tanh(\DeltaEPlusMinus/2kT)/\VQPClr\DeltaEPlusMinus]/\DeltaEPlusMinus^{3}\cosh^{2}(\DeltaEPlusMinus/kT)$.
Note that the second term in Eq.~(\ref{eqnCurrentStat}) depends
on $\DE$, a property which can be used for read out, as we have discussed
above. For example, for different Zeeman splittings and $\DE_{\uparrow,\downarrow}=\pm\Dm/2$,
$\decohRLintrinsic=10^{9}\:\mathrm{s^{-1}}$, $I_{\ixLeftDot}=1\,\mathrm{nA}$,
and $I_{\ixRightDot}=0$, the current difference is $\currentStat_{\downarrow}-\currentStat_{\uparrow}=0.4\,\mathrm{nA}$,
which reduces to $0.05\,\mathrm{nA}$ for $I_{\ixRightDot}=0.5\,\mathrm{nA}$.
However, typical QPC currents currently reachable are $I_{\ixLeftDot}=10\,\mathrm{nA}$
and $I_{\ixRightDot}=9.9\,\mathrm{nA}$, i.e., the relaxation of the
double dot due to the QPC is suppressed, $\eta<10^{-3}$, and other
relaxation channels become important.

\emph{Incoherent tunneling}. So far, we have discussed coherent tunneling.
We can also take incoherent tunneling into account, e.g., phonon assisted
tunneling, by introducing relaxation rates in Eqs.\ (\ref{eqnMasterL}),(\ref{eqnMasterRL}).
For example, for detailed balance rates and neglecting coherent tunneling,
we find the stationary current $\currentStatIncoh=\frac{1}{2}(I_{\ixLeftDot}+I_{\ixRightDot})+\frac{1}{2}(I_{\ixRightDot}-I_{\ixLeftDot})\tanh(\DE/2kT)$
(which becomes $I_{\ixRightDot}$ for $\DE>kT$). The QPC current
again depends on $\DE$ and can be used for spin read out. The current
can also be measured on shorter time scales as we discuss now. 

\emph{Read out with time-dependent currents} is possible if there
is sufficient time to distinguish $I_{\ixLeftDot}$ from $I_{\ixRightDot}$
between two tunneling events to or from the reference dot, i.e., we
consider $\decohRLtot>\tDD$. In this incoherent regime, the tunneling
from qubit to reference dot occurs with a rate $\chargeDecay{\uparrow}$
or $\chargeDecay{\downarrow}$, depending on the qubit state, with,
say, $\chargeDecay{\downarrow}\ll\chargeDecay{\uparrow}$. Such rates
arise from spin-dependent tunneling, $\tDD^{\uparrow,\downarrow}$,
or from different Zeeman splittings and tuning to tunneling resonance
for, say, qubit $\spup$ while qubit $\spdown$ is off-resonant, see
Figs.~\ref{figROdd}a and \ref{figROdd}b. For read out, the electron
is initially on the left dot and the QPC current is $I_{\ixLeftDot}$.
Then, if the electron tunnels onto the reference dot within time $t$
and thus changes the QPC current to $I_{\ixRightDot}$, such a change
would be interpreted as qubit in state $\spup$, otherwise as qubit
$\spdown$. For calculating the measurement efficiency $\efficiencyMeas$,
we note that $p_{\uparrow}=p_{0}=1-e^{-t\chargeDecay{\uparrow}}$
and $p_{\downarrow}=p_{1}=e^{-t\chargeDecay{\downarrow}}$ (with this
type of read out, $\chargeDecay{\downarrow}$ corresponds to a loss
of the information, i.e., describes {}``mixing'' \cite{Schoen}).
We then maximize $\efficiencyMeas$ by choosing a suitable $t$ and
find efficiencies $\efficiencyMeas\gtrsim50\,\%$ for $\chargeDecay{\uparrow}/\chargeDecay{\downarrow}\gtrsim8.75$
and $\efficiencyMeas\gtrsim90\,\%$ for $\chargeDecay{\uparrow}/\chargeDecay{\downarrow}\gtrsim80$.

A more involved read out is to measure the current through the QPC
at different times. The current as function of time switches between
the values $I_{\ixLeftDot}$ and $I_{\ixRightDot}$, i.e., shows telegraph
noise, as sketched in Fig. \ref{figTelegraph}d. Since the frequency
of these switching events (roughly $\chargeDecay{\uparrow}$ or $\chargeDecay{\downarrow}$)
depends on the spin, the QPC noise reveals the state of the qubit.
Finally, at times of the order of the spin relaxation time $T_{1}$,
the information about the qubit is lost. At each spin flip, the switching
frequency changes ($\chargeDecay{\uparrow}\leftrightarrow\chargeDecay{\downarrow}$),
which thus provides a way to measure $T_{1}$.

In conclusion, we have given the criterion when $n$-shot measurements
are possible and have introduced the measurement efficiency $\efficiencyMeas$.
For electron spin qubits, we have proposed several read-out schemes
and have found efficiencies up to 100\%, which allow single-shot read
out. Other schemes, which are based on the Pauli principle, have a
lower efficiency, $\efficiencyMeas=50\%$. We thank Ch.~Leuenberger
and F. Meier for discussions. We acknowledge support from the Swiss
NSF, NCCR Nanoscience Basel, DARPA, and ARO.

\clearpage


\begin{thebibliography}{10}
\bibitem{fnUncertOutcome}In other words, we assume a sufficient signal-to-noise ratio of the
apparatus to distinguish the measurement outcome $\measOutcome{0}$
from $\measOutcome{1}$.
\bibitem{effSuperpos}For a qubit in an arbitrary superposition $\alpha\ket{0}+\beta\ket{1}$,
the expectation value of the measurement is $\langle\hat{A}\rangle=|\alpha|^{2}\expectQubitMeas{\hat{A}}{0}+|\beta|^{2}\expectQubitMeas{\hat{A}}{1}$,
which allows to determine $|\alpha|^{2}$ and $|\beta|^{2}=1-|\alpha|^{2}$.
(To measure the phase $\arg\alpha/\beta$, first some single qubit
rotations need to be performed.) In order to differentiate a given
$|\alpha|^{2}$ from a value $|\alpha'|^{2}$, a sufficient $n$ is
given by Eqs.~(\ref{eqnnMin}) and (\ref{eqnMEffG}) after replacing
$p_{0}\to|\alpha|^{2}p_{0}+(1-|\alpha|^{2})(1-p_{1})$ and $p_{1}\to1-|\alpha'|^{2}p_{0}-(1-|\alpha'|^{2})(1-p_{1})$.
\bibitem{footnoteExpG}If $n$ is small, one can use Clopper-Pearson confidence intervals.
However, if read out of one state is perfect, say $p_{0}=1$, we can
no longer approximate with a normal distribution, even for large $n$.
In that case, finding $\measOutcome{0}$ as outcome $n$ times in
a row, even if the qubit is $\ket{1}$, i.e., read out fails, occurs
with probability $(1-p_{1})^{n}$. Thus, $n\geq\log(\alpha)/\log(1-\efficiencyMeas)$
is sufficient for read out.
\bibitem{footnoteRefLead}Instead of a reference dot, the qubit dot can be coupled to a lead.
To ensure that only electrons with, say, spin $\downarrow$ can tunnel,
one can use spin-polarized leads or a Zeeman splitting on the dot
and properly tuned energy levels \cite{ELesr}.
\bibitem{footnoteTwoLevel}We map the two-level system $\{\ket{\lblL},\,\ket{\lblR}\}$ onto
a pseudo spin $\frac{1}{2}$ with Hamiltonian $H(t)=\frac{1}{2}\DE\sigma_{z}+\tDD\sigma_{x}+X(t)\sigma_{z}+\frac{1}{2}(\tQPC_{\ixLeftDot}+\tQPC_{\ixRightDot})V_{\mathrm{QPC}}(t)$.
The fluctuations due to the QPC are $X(t)=\frac{1}{2}(\tQPC_{\ixLeftDot}-\tQPC_{\ixRightDot})V_{\mathrm{QPC}}(t)$
with $V_{\mathrm{QPC}}(t)=e^{iH_{\mathrm{\ixQPCleads}}t}V_{\mathrm{QPC}}e^{-iH_{\mathrm{\ixQPCleads}}t}$.
\bibitem{footnoteBloch}We define $\BlochEqM=\frac{1}{2}\mathrm{Tr}\,\rho\,\bsigma$ and write
the master equation in the standard Bloch notation, $\dot{\BlochEqM}=\bomega\times\BlochEqM+\BlochIncohMatrix'(\BlochEqM-\BlochEqM_{0})$,
with $\bomega=\left(2\tDD,0,0\right)$ and $\BlochIncohMatrix'\vec{a}=(0,\;-\tDD\VQPClr/\Dm-\tDD\VQPClr(\fermiProdOverDmPMSum-2\fermiProdOverDmSymbol_{0})\DE/\DeltaEPlusMinus^{2},\;\DE+\tDD\VQPClr/\Dm)\times\vec{a}+\BlochIncohMatrix\vec{a}$,
where $\BlochIncohMatrix$ is symmetric with elements $\BlochIncohMatrix_{xy}=\BlochIncohMatrix_{yx}=-\tDD\VQPClr/\Dm$;
$\BlochIncohMatrix_{xz}=\BlochIncohMatrix_{zx}=\BlochIncohMatrix_{xy}-\tDD\VQPClr(\fermiProdOverDmPMSum-2\fermiProdOverDmSymbol_{0})\DE/\DeltaEPlusMinus^{2}$;
$\BlochIncohMatrix_{yy}=-\VQPClr(1+\fermiProdOverDmPMSum)+\VQPClr(\fermiProdOverDmPMSum-2\fermiProdOverDmSymbol_{0})\DE^{2}/\DeltaEPlusMinus^{2}$;
$\BlochIncohMatrix_{xx}=2\BlochIncohMatrix_{xy}+\BlochIncohMatrix_{yy}$;
$\BlochIncohMatrix_{yz}=\BlochIncohMatrix_{zy}=\BlochIncohMatrix_{zz}=0$.
Finally, $\BlochEqMSymbol_{0x}=\BlochEqMSymbol_{0y}=0$ and $\BlochEqMSymbol_{0z}=[\tDD\VQPClr(2\tDD+\DE)/(1+\fermiProdOverDmPMSum)\Dm^{2}-\BlochIncohMatrix_{xy}]/\BlochIncohMatrix_{xz}$.
\end{thebibliography}
\end{document}